\documentstyle[prl,aps,epsf]{revtex}\def\narrowtext{}\tighten\twocolumn
\begin{document}
\draft

\title{Electronic structure of MgB$_2$ from angle-resolved photoemission 
spectroscopy}
\author{H. Uchiyama$^{1}$, K.M. Shen$^{2}$, S. Lee$^{1}$, A. Damascelli$^{2}$, D.H. Lu$^{2}$, D.L. Feng$^{2}$, 
Z.-X. Shen$^{2}$ and S.Tajima$^{1}$}

\address{$^{1}$Superconductivity Research Laboratory, International Superconductivity 
Technology Center, 1-10-13 Shinonome, Koto-ku, Tokyo 135-0062, 
Japan\\
$^{2}$Department of Physics, Applied Physics and Stanford Synchrotron 
Radiation Laboratory, Stanford University, Stanford, CA, 94305, 
USA.}


\address{\begin{minipage}[t]{6.0in} %
\begin{abstract}%
The first angle-resolved photoemission spectroscopy results from MgB$_2$ 
single crystals are reported. Close to the $\Gamma$K and $\Gamma$M directions, three distinct
dispersions are observed approaching the Fermi energy, as
can be assigned to the theoretically predicted $\sigma$ (B $2p_{x,y}$) and $\pi$ (B $2p_z$) bands. In addition, we also observed
a small parabolic-like band centered around $\Gamma$, and attributed it to a
surface-derived state.  Good agreement between
our results and the band calculations suggests that the electronic structure of MgB$_2$ is of a conventional
nature, thus implying that electron correlations are weak, and may be of little importance to the
superconductivity in this system. 
\typeout{polish abstract} %
\end{abstract}
\end{minipage}} %

\maketitle %
\narrowtext
The discovery of superconductivity in MgB$_{2}$~\cite{disc} has garnered a 
tremendous amount of interest, particularly regarding whether this material can be explained
within a conventional BCS framework~\cite{BCS}. The transition temperature ($T_c$)  of 39 K is considerably
higher than what many would have originally believed possible from a conventional electron-phonon
interaction alone~\cite{mcmillan,Allen}, and while isotope substitution shows that this compound can be fairly well explained
by considering an electron-phonon interaction~\cite{iso1,iso2}, electron correlations may additionally conspire to raise
$T_c$~\cite{Hirsch,Imada,Furukawa}.

In order to address the degree of electronic correlations in MgB$_2$, an experimental determination of its electronic structure
is crucial. As is well known, the strong Coulomb repulsion in high-$T_c$ cuprates and
other correlated materials, can dramatically affect the low-energy excitations~\cite{Mott}. While
calculations of the band structure have already been carried out~\cite{band1,band2}, there has been, to date, no experimental
data to corroborate these theoretical predictions. 

Angle-resolved photoemission spectroscopy (ARPES) is uniquely powerful in its capability to directly probe the electronic
structure of the solid state. Particularly for the cuprates, this technique has played a pivotal role in shaping  our current
understanding of their electronic structure as well  as their electron-electron correlations~\cite{shen,Olson}. In this letter,
with the recently  successful synthesis of small single crystals~\cite{Lee}, we can, for the first time, directly observe its
electronic  structure of MgB$_{2}$ by ARPES, thereby determining if MgB$_{2}$ is simply an ordinary 
metal with optimized superconducting parameters, or whether other effects such as electron correlations play a role in
mediating  the superconductivity.


Single crystals of MgB$_{2}$ were grown in the quasi-ternary Mg-MgB$_{2}$-BN system under 5-6 GPa at
1600 $^\circ$C. Several single  crystals with typical dimensions of 0.3$\times$0.3$\times$0.1 mm$^{3}$
were  selected for this study. Single-crystallinity was confirmed 
by four circle XRD. Both resistivity and magnetization measurements 
verified that the crystals exhibit superconductivity at 38 K 
with a sharp transition width of 0.3 K~\cite{Lee}. The ARPES measurements 
were performed at Beamline 5-4 of the Stanford Synchrotron Radiation 
Laboratory with a total energy resolution of better than 40 meV 
and an angular resolution of $\pm$0.15$^\circ$. In this 
case, higher energy resolution was sacrificed in order to obtain 
reasonable counting statistics on the small samples. The samples 
were first aligned by Laue diffraction, then cleaved {\it in situ }
in the $ab$ plane at a pressure of better than 5$\times$10$^{-11}$ torr 
and below 10 K. Due to the observed rapid degradation of the 
sample surfaces, all measurements were taken within 5 hours of 
cleaving. Reproducible results were observed on different cleaves.

ARPES spectra were collected parallel to the high symmetry directions, $\Gamma$=(0,0,0)M=($\pi$,0,0)
and $\Gamma$K=(2/3$\pi$,2/3$\pi$,0), using  a photon energy of $h\nu$ = 28 eV with the polarization 
perpendicular to each respective symmetry direction (see Figure 1a). For the former direction, polar emission angle was
changed from 0$^{\circ}$ (normal emission) to 28$^{\circ}$ toward [100]-axis, while for the latter it was changed from
0$^{\circ}$ to 33$^{\circ}$ toward [110]-axis. Since MgB$_{2}$ is characterized by a three dimensional  electronic structure,
and the band dispersion along the
$k_z$-axis  is predicted to change dramatically~\cite{band1,band2}, we performed photon-energy dependence measurements at normal
emission in order to estimate  the corresponding
$k_z$ value for each given incident photon  energy~\cite{hufner}. The photoemission cross section was found 
to be maximum at $h\nu$= 28 eV between 17 and 28 
eV, and decreased quickly and monotonically when going to progressively 
lower photon energies. Due to the
combination of the large background and low intensity, it was not
possible to observe clear change 
of the electronic bands at normal emission as a function of photon energy. Therefore, we could not determine exact
$k_z$-coordinate experimentally.

The spectra in Figures 1b and 1c are representative energy distribution 
curves (EDCs) taken along the $\Gamma$(A)-K(H) and $\Gamma$(A)-M(L) 
directions, respectively, and show the presence of multiple dispersive 
bands as marked by the colored dots. The positions of these markers 
were determined from the second derivative plots, as will be 
described in the next paragraph. Near the respective K(H) and M(L)
points, we observe a strong peak whose dispersion approaches and crosses the Fermi energy ($E_F$), 
as denoted by the blue dots. Along $\Gamma$(A)-K(H), another feature, 
marked in green, is observed approaching $E_F$ near the $\Gamma$(A)
point. However, along the $\Gamma$(A)-M(L) direction, the corresponding 
feature is very weak, although subtle changes in the lineshape 
can be discerned, and are marked in green. Finally, along both 
the $\Gamma$(A)-K(H) and $\Gamma$(A)-M(L)  directions, a small parabolic-like 
band is centered near $\Gamma$(A) and crosses $E_F$ near ($\pi$/4, 
0, $k_z$) and ($\pi$/6,$\pi$/6, $k_z$) respectively.

To more effectively visualize the ARPES data in the context of 
band dispersions, image plots of the second derivative of the 
EDCs are shown in Figure 2a. By taking the second derivative 
of the raw data in Figures 1b and 1c, the relative contrast of existing features can be 
enhanced. We note that spurious intensities in the second derivative plots (due to the sensitivity to statistical
noise) which do not appear in the raw data itself, likely do not represent any true feature. The features identified
in Figures 1b and 1c are
 evident in these image plots and the overlaid solid lines are  the experimental band dispersions. The two weak
features in the  raw data along the $\Gamma$(A)-M(L) direction are more pronounced 
in the second derivative image plots. A comparison of these experimentally 
observed features with the band structure calculations for $k_z \approx 0$ is shown 
in Figure 2b, and exhibits remarkable agreement between the experiment and theory~\cite{band1,band2}. 
To account for the uncertainties in $k_z$,  we compare 
our results to band structure calculations taken along the $\Gamma$M 
and $\Gamma$K directions, but projected in $k_z$  from $k_z$=0 
to $k_z$=$\pm 0.14\pi$, that are shown by broad 
lines. Since this agreement was completely lost in other theoretical $k_z$ regions, we conclude that for
normal emission at
$h\nu$=28 eV we are close to $\Gamma$, and thus our angular cuts lay close to the $\Gamma$M and $\Gamma$K lines. From Figure
2b, we can clearly assign the higher intensity  feature, marked in blue, to the boron 2$p_z$ ($\pi$)  band, and the lower
intensity features, marked in green, to the boron  2$p_{x,y}$ ($\sigma$) bands. Although a single
$\sigma$-band  is observed along $\Gamma$K, the band structure calculations 
predict that two 2$p_{x,y}$ bands lie close together in 
energy, and therefore the broad feature we observe likely results 
from the superposition of the two. In addition, while the 2$p_{x,y}$ 
bands along $\Gamma$M are weak in both the EDCs and even in 
the second derivative image plots, the close agreement of these 
features with the theoretical calculations lends strong support 
to our identification of these bands. We should also note that 
a similar contrast in the photoemission matrix element intensity 
between the $\sigma$ and $\pi$ bands is observed in graphite~\cite{graphite}, 
which possesses a somewhat similar electronic structure to MgB$_{2}$.

Finally, we turn our attention to the state centered around $\Gamma$
and denoted in red, as shown in Figure 3. From comparisons with 
the band structure, there is no theoretically predicted bulk 
band which would correspond to this particular feature. However, 
as ARPES is extremely surface sensitive, it is entirely probable 
that this feature originates from a surface electronic state. The 
existence of surface states, which arise from the broken translational 
symmetry at the crystal surface, is a rather universal phenomenon 
and occurs in many simple materials such as Au, Ag, Cu, Si, and 
graphite~\cite{hufner}. Therefore, the existence of such a surface state 
in MgB$_{2}$ should not be surprising, especially around the $\Gamma$
point, where there is a gap in the $k_z$-projected bulk 
band structure. The close agreement of all other features to 
the band structure calculations also lends credence to our assignment 
of this feature as a surface state, and not some unpredicted 
bulk band. In fact, recent calculations
including the effect of surface termination find that this observed state in Figure 3 agrees well with a predicted
surface-derived band on Mg-terminated surface layers, both in terms of its location in momentum and energy, as well as its
dispersion\cite{private}, thus well explaining the origin of this feature. It should be noted that many measurements of the
gap  magnitude have been conducted by surface-sensitive techniques  such as tunnelling~\cite{tunnel1,tunnel2,tunnel3,tunnel4}
and angle-integrated photoemission~\cite{PES1,PES2},  and the presence of a surface state of unknown origin could potentially 
complicate the interpretations of these measurements, especially  those which discuss a multiple gap feature.

The close overall agreement of our experimentally determined 
band dispersions with the theoretical calculations seems to indicate 
that the effects of electron-electron correlations in this material 
are surprisingly weak. Because the electron-electron correlations 
persist to high energies, this results in an overall renormalization, 
or narrowing, of the total bandwidth. Clearly, this behavior 
is not observed, even up to binding energies as high as 2.5 eV, 
thus suggesting that the correlations in this material are fairly 
unimportant and that MgB$_2$ can be well-described by band theory. 
However, at lower energy scales, it is often possible to observe the 
coupling of the quasiparticles to collective modes such as phonons~\cite{Be}, 
as might be expected in a relatively strong-coupling superconductor 
such as MgB$_2$ where the phonon DOS extends out to 100 meV~\cite{neut}. 
This was not observed in our measurements, primarily, we believe, 
due to the broadness of the measured lineshapes. Nevertheless, 
our clear measurements of the overall electronic structure tend 
to support the assumption that MgB$_2$ is a conventional metal 
and that the mechanism for superconductivity is likely to be 
of a conventional origin. 

 We are grateful to J. Kortus and I.I. Mazin 
for kindly providing their band structure calculations and useful suggestions, and to S. Drechsler and V. Servedio for showing
their unpublished results. This  work was supported by the New Energy and Industrial Technology 
Department Organization (NEDO) as Collaborative Research and 
Development of Fundamental Technologies for Superconductivity 
Applications. SSRL is operated by the DOE Office of Basic Energy 
Research, Division of Chemical Sciences. The office's division 
of Material Science provided support for this research. The work 
at Stanford University was also supported by NSF Grant No. DMR0071897 and 
ONR Grant No. N00014-98-1-0195.

\begin{figure}
\caption{ a) Brillouin zone of MgB$_{2}$, with the speculated trajectories of our measurements through the zone shown in
purple and black. For the details, see the text. b) Selected energy distribution curves (EDCs) taken along the
$\Gamma$(A)-K(H) direction (purple line). The B $2p_z$  band is marked in blue, and the B $2p_{x,y}$ feature is 
marked in green. Red dots denote the probable surface state feature 
centered around $\Gamma$(A). c) Selected EDCs taken along the $\Gamma$(A)-M(L) direction (black line). The B $2p_z$  band is
again marked in blue,  and the possible B $2p_{x,y}$ bands are marked in green. 
Red dots denote the probably surface state feature centered around $\Gamma$(A).}
\label{rough}
\end{figure}

\begin{figure}
\caption{ a) Second-derivative plots of the EDCs taken along the 
cuts shown in Figure 1. Data was smoothed in both energy and 
momentum before taking the second derivative. Colored lines are 
experimentally measured band dispersions as determined by eye, 
with the color coding consistent with Figures 1b and 1c. b) Comparison of experiment and theory. Experimentally determined 
dispersions are shown in color, and theoretically predicted dispersions 
shown in black. The width of the theoretical lines represents 
the projection of $k_z$ values from 0 to $\pm 0.14\pi$.}
\label{der}
\end{figure}

\begin{figure}
\caption{ Spectra taken close to $\Gamma$ along the $\Gamma$K 
direction which show the probable surface state (red) and the B
2$p_{x,y}$ band (green)}
\label{surf}
\end{figure}

\end{document}